\documentclass[journal]{IEEEtran}

\usepackage{physics} 
\usepackage{textcomp} 
\usepackage{siunitx} 
\usepackage{enumerate} 
\usepackage{pgfplots}
\usepackage{pgfplotstable}
\usepackage{tikz,pgfplots}
\usepackage{float}
\usepackage{multicol}
\usepackage{graphicx}
\usepackage{cite}
\usepackage{hyperref}
\usepackage{amsmath}
\usepackage{amssymb}
\usepackage{wasysym} 
\usepackage{booktabs}
\usepackage{amsmath,amsfonts}
\usepackage{algorithmic}
\usepackage{array}
\usepackage[caption=false,font=normalsize,labelfont=sf,textfont=sf]{subfig}
\usepackage{stfloats}
\usepackage{url}
\usepackage{verbatim}
\def\BibTeX{{\rm B\kern-.05em{\sc i\kern-.025em b}\kern-.08em
    T\kern-.1667em\lower.7ex\hbox{E}\kern-.125emX}}
\usepackage{balance}
\usepackage{soul}
\usepackage{makecell}
\usepackage{adjustbox}

\usepackage[normalem]{ulem}

\AtBeginDocument{\RenewCommandCopy\qty\SI}

\title{Accurate Modeling of Directional Couplers with Oxide Cladding: Bridging Simulation and Experiment}

\author{
    \IEEEauthorblockN{Yuval Warshavsky\IEEEauthorrefmark{1}\IEEEauthorrefmark{3}, Yehonathan Drori\IEEEauthorrefmark{2}, Jonatan Piasetzky\IEEEauthorrefmark{1}, Amit Rotem\IEEEauthorrefmark{2}, Ofer Shapiro\IEEEauthorrefmark{2}, Yaron Oz\IEEEauthorrefmark{1}, Haim Suchowski\IEEEauthorrefmark{1}\\ }
    \IEEEauthorblockA{\IEEEauthorrefmark{1}Raymond and Beverly Sackler School of Physics and Astronomy, Tel Aviv University, Tel Aviv, 6997801, Israel\\ }
    \IEEEauthorblockA{\IEEEauthorrefmark{2}Quantum Pulse Inc.\\ } \IEEEauthorblockA{\IEEEauthorrefmark{3}warshavsky1@mail.tau.ac.il}
}

\begin{document}

\maketitle

\begin{abstract}
    Directional couplers are a fundamental building block in integrated photonics, particularly in quantum applications and optimization-based design where precision is critical. Accurate functionality is crucial to ensure reliable operation within classical and quantum circuits. However, discrepancies between simulations and measurements are frequently observed. These inaccuracies can compromise the performance and scalability of integrated photonic systems, underscoring the critical need for advanced, precise simulation methods that bridge the gap between design and implementation. In this work, we show that this discrepancy can be mainly attributed to density changes in the oxide cladding. We conduct a systematic study involving experimental optical measurements, numerical simulations, and direct electron microscopy imaging to investigate this discrepancy in directional couplers. We find that the impact of cladding density variations on performance increases as feature gaps shrink. By incorporating these effects into our simulations using a novel and physically motivated \textit{Effective Trench Medium Model} (ETMM), we achieve highly accurate reproduction of experimental measurements. We quantify the effects of cladding density variations on the SU(2) symmetry parameters that govern light propagation in directional couplers. This insight is crucial for advancing the precision of compact device fabrication, enabling reliable simulation of photonic integrated devices.
    
    

\end{abstract}


\section{Introduction}

\IEEEPARstart{I}{ntegrated} photonics has emerged as a transformative technology, revolutionizing fields such as telecommunications, quantum information processing (QIP), sensing, and machine learning \cite{6146487, Wang2020, siew2021review, sunny2021survey}. The integration of photonic components on a single chip promises unprecedented levels of functionality and performance while simultaneously reducing size, cost, and power consumption. Silicon photonics, in particular, has garnered significant attention due to its compatibility with mature CMOS fabrication techniques, enabling cost-effective and precise nanoscale fabrication, as well as seamless integration with nanoelectronics \cite{4032660, shekhar2024roadmapping}.

At the heart of photonic integrated circuits (PICs) lie directional couplers (DCs), fundamental building blocks crucial for power splitting, interference, mode separation, and recently as a photonic quantum gate in QIP \cite{knill2001scheme, kok2007linear, o2007optical, yariv2007photonics, chrostowski2015silicon, lifante_integrated_2003, Wang2020, okamoto2010fundamentals}. Light propagation in these devices follows SU(2) symmetry, making them particularly well-suited for implementing single-qubit gates within the dual-rail encoding scheme. 

Despite their ubiquity and importance, accurate simulation and prediction of DC performance remains challenging, with significant discrepancies often observed between simulated and measured results. These inconsistencies have historically been attributed to unknown errors or fabrication variations, leading to a reliance on "trial and error" approaches in DC design and optimization \cite{chrostowski2015silicon, bogaerts2018silicon}. Such empirical methodologies hamper efficient device development and limit the potential for advanced, optimization-based designs in more complex photonic structures.

\begin{figure}[!t]
    \centering
    \includegraphics[width=1\linewidth]{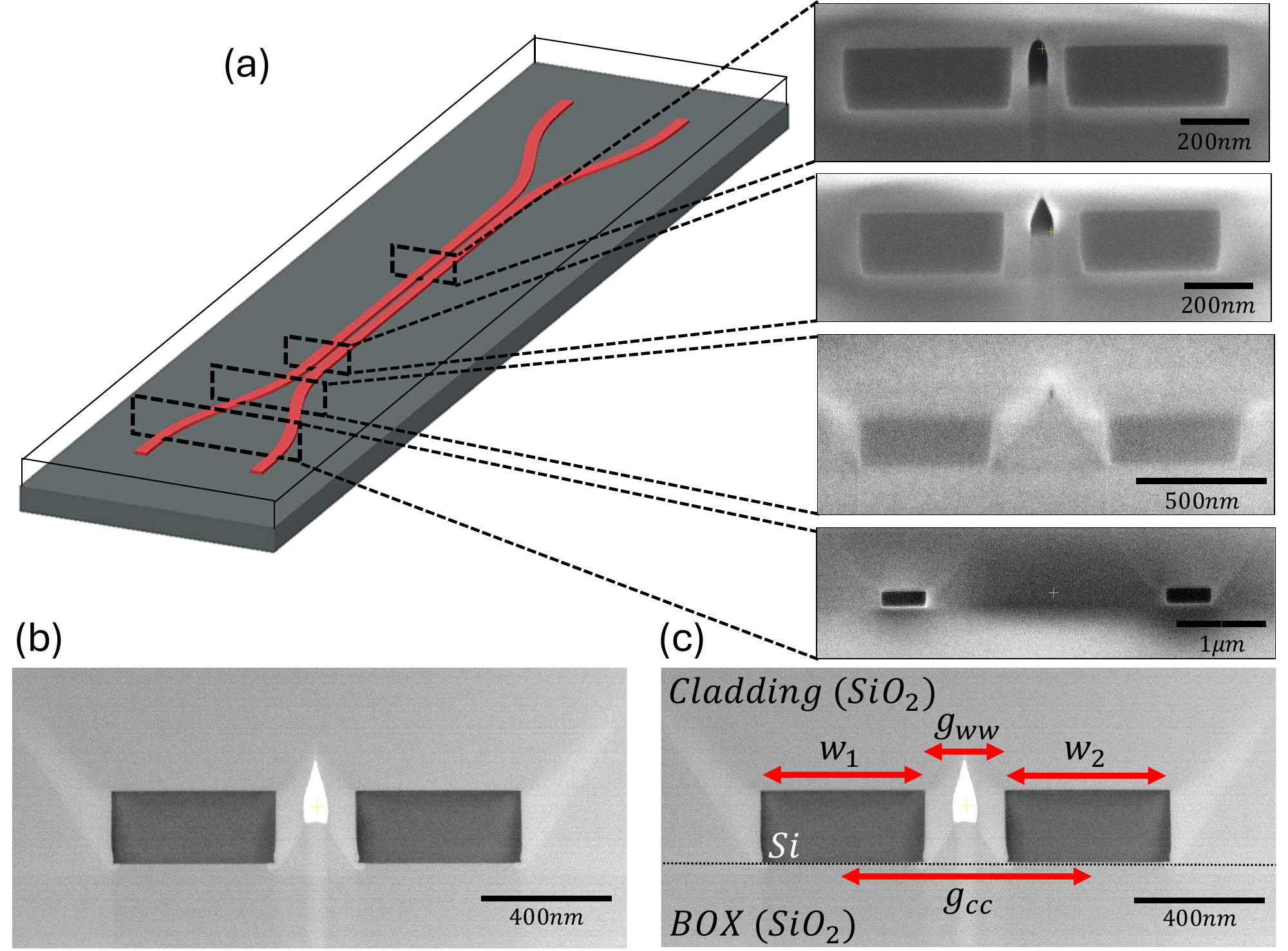}
    \caption{SEM cross-section images of a SOI directional coupler with oxide cladding for various gaps. \textbf{(a)} Perspective view of a directional coupler, along with SEM images of various cross-sections, each with a different value of wall-to-wall gap ($g_{ww}$). As the waveguides get closer and $g_{ww}$ decreases, a void and cladding inhomogeneities appear. \textbf{(b)} High-resolution STEM cross-section image of a directional coupler. Clearly shown is a void between the waveguides, and the cladding inhomogeneities in the shape of "wings". \textbf{(c)} Geometrical dimensions of the directional coupler and the materials. $g_{ww}$ is wall to wall gap, $g_{cc}$ is center to center gap and $w_i$ is the width of waveguide $i$. The layer under the line is the \textit{Buried Oxide} (BOX) layer, manufactured separately from the cladding.}
    \label{fig:dc_and_sem}
\end{figure}

Several observations of the discrepancy have been made \cite{Tseng:13, emre2019experimental, mikkelsen2014dimensional}. Noticeably Tseng et al. and Mikkelsen et al. experimentally show deviations that increase with small gaps in directionally coupled devices. However, a complete map between simulations and experimental observation is still missing.

Here, we present the root cause of discrepancies in the performance of directional couplers covered with an industry-standard oxide cladding. Through systematic analysis and advanced characterization techniques, we demonstrate that the presence of density differences in oxide cladding - a phenomenon previously reported but not fully studied in the context of integrated photonics
\cite{pung2012fabrication, 9252413,halir2016ultra, ohkubo2003fabrication, schwartz1992gap, logan1989study, riley1989limitation}  -
is the primary factor responsible for deviations from the simulated behavior. 

We have utilized electron microscopy imaging to show the formation of voids in the cladding medium in the trench between the waveguides (see Fig. \ref{fig:dc_and_sem}).
We demonstrate that the discrepancies between measurement and simulation cannot be attributed solely to these voids. Instead, they arise from density variations that include the voids within the medium.
Moreover, we show that using a novel \textit{Effective Trench Medium Model}, simulations recover measured results with unprecedented accuracy.
Last, we show experimentally that these density changes are only dependent on the wall-to-wall gap between waveguides, and are consistent throughout different devices on chips from different fabrication cycles and can be accounted for a priori.

Our comprehensive analysis enables accurate simulation of directional couplers and, by extension, a wide range of integrated photonic devices. Our findings are particularly significant for compact devices where the influence of voids and density variation are significant.

This paper is organized as follows: we first present the theory for light propagation in directional couplers, with the general SU(2) parameters that govern it. Next, we discuss the fabrication process and its role in the discrepancy between simulation and measurement.
We then show that including the void in simulations yields better results than previous models but still insufficient in explaining the discrepancies.
Last, we present the ETMM, show that it accurately predicts measurements, and discuss its implications on PIC design.

\section{Framework for Directional Couplers}

\noindent Directional couplers are integrated devices in which two waveguides are brought closely together to enable optical energy transfer between them by evanescent coupling (see Fig.~\ref{fig:dc_and_sem}a).  
For single-mode waveguides, the evolution of the electric fields in the two waveguides is described by coupled mode theory (CMT) \cite{yariv2007photonics, okamoto2010fundamentals}:

\begin{equation}
    \begin{pmatrix}
        E_1(L) \\ E_2(L)
    \end{pmatrix}
    = U
    \begin{pmatrix}
         E_1(0) \\ E_2(0)
    \end{pmatrix},
    \label{eq:unitary propagator on fields}
\end{equation}
where the unitary propagator is defined as:

\begin{equation}
\begin{split}
         & U (\kappa, \Delta\beta,L)= \\
         & \begin{pmatrix}
\cos(s L) - i\frac{\Delta \beta}{s}\sin(s L) & - i\frac{\kappa}{s}\sin(s L) \\ 
- i\frac{\kappa}{s}\sin(s L) & \cos(s L) + i\frac{\Delta \beta}{s}\sin(s L).
\end{pmatrix}
\end{split}
\label{eq:unitary propagator}
\end{equation}

Here $\kappa$ is the coupling coefficient between two propagating modes, which stems from the overlap of one mode's evanescent tail with the mode in the other waveguide.
$\Delta \beta = (\beta_1-\beta_2)/2$ is the propagation constant difference between the two waveguides affecting the inversion length and the maximal power transfer of the DC.
$L$ is the interaction length, defining the region where waveguides are coupled and energy is exchanged. We also define the effective coupling parameter:

\begin{equation}
\label{eq:rabi frequency}
s = \sqrt{\kappa^2 + \Delta \beta^2} 
\end{equation}

The propagator in Eq.~\ref{eq:unitary propagator} exhibits SU(2) symmetry, such that the power of light propagating in a DC oscillates between the DC ports with spatial frequency $s$ as a function of $L$, analogous to temporal Rabi oscillations in two-level systems in atomic physics \cite{mrejen2015adiabatic}.


The coupling coefficient $\kappa$ is the primary property of a directional coupler and dictates the amount of energy transferred from one waveguide to the other per unit length. 
The coupling magnitude decays exponentially with the distance between the centers of the waveguide, a quantity denoted by $g_{cc}$ (center to center gap, see Fig.~\ref{fig:coupling}a), as it's proportional to the overlap between the modes. 
 
The splitting ratio (SR), the fraction of the power transferred between the waveguides, is given by: 

\begin{equation}
SR = |U_{12}|^2 = |U_{21}|^2 = \frac{\kappa^2}{s^2}\sin^2\left[s\left(L+L_{S bend}\right)\right]
\label{eq:SR}
\end{equation}

Where $L_{S bend}$ accounts for the propagation of light in the regions where the DC ports approach or depart from each other, called the S Bend \cite{chrostowski2015silicon}. In the case of symmetric DCs, with identical cross-sectional geometries where $w_1=w_2$ and consequently $\Delta\beta=0$, the SR can fully characterize the operation of the DC in terms of amplitude and phase as $\kappa=s$. Only in symmetric DCs is complete population transfer (CPT) possible.


\section{Fabrication Process}


\noindent The devices are fabricated in an industry-standard silicon-on-insulator (SOI) fabrication process, with a silicon thickness of 220\,nm on top of a buried-oxide (BOX, $SiO_2$) layer performed by \textit{Applied Nanotools inc.} (ANT).
The fabrication process includes four steps: First, the silicon layer is patterned with a mask material using electron beam lithography (EBL). Second, the silicon layer is etched with an anisotropic inductively coupled plasma-reactive ion etching (ICP-RIE) process down to the BOX layer. Third, the mask material is chemically removed. Last, the etched silicon layer is covered with an oxide cladding using a plasma-enhanced chemical vapor deposition (PECVD) process. This step is crucial to protect the microscopic etched patterns and enable the placement of wiring and metal elements for active control.

The PECVD process is the reason for the creation of the voids and density variations. The $SiO_2$ hardens amorphously in the high aspect ratio trenches between waveguides, creating the inhomogeneities \cite{halir2016ultra, pung2012fabrication, ohkubo2003fabrication, schwartz1992gap, logan1989study, chang2004trench}, shown in Fig.~\ref{fig:dc_and_sem}.

\section{Measurement and Simulation of Directional Couplers}


\noindent We have measured the splitting ratio of 117 unique DC cross-sections over two chips from different fabrication cycles, as described in Table \ref{tab:1}. To accurately determine the SR of each DC, light was launched via a fiber array and integrated grating couplers into the DC. By measuring both input and output ports for each DC (four measurements total), we effectively eliminated the influence of insertion loss variations, including variations both in the grating couplers' efficiency and fiber-chip alignment. This method, detailed in Ref. \cite{emre2019experimental}, avoids the need for insertion loss calibration, leading to significantly reduced measurement errors. The remaining significant variations in the measured SRs are due to fabrication errors of the DCs themselves.
Of the 117 cross-section designs, 55 were placed on a single chip and split into calibration and validation sets, containing 13 and 42 cross-section geometries, respectively. Each cross-section design in the calibration set contains 7 different interaction lengths, and each in the validation set contains 9 lengths, totaling 469 DC designs. Moreover, each design was fabricated four times and randomly positioned on the chip to reduce locally correlated fabrication errors, totaling 1876 DC instances measured on a single chip. The final SR measurement of that design is the average of the 4 instances measured with the corresponding statistical uncertainty. 
Of the 117 different cross-section geometries, the remaining 62 were placed on a chip from a different fabrication cycle, each with 8 different interaction lengths, totaling 496 designs. On this chip, there is only one instance of each DC. These DCs serve as an additional validation set.

To rapidly and reliably measure the entire set of 1876 DCs on a single chip, automation was implemented. For each DC, the chip is moved using a motorized stage to coarsely position the DC under the fiber array. Then, piezo actuators, driven by a feedback loop, were used to fine-tune the fiber to DC coupling.

Several DCs with varying $L$ were designed for each set of waveguide widths and gap and placed on the chip. By fitting Eq.~\ref{eq:SR} to the measured SR as a function of $L$, we determine the frequency $s$. Comparing the extracted $s$ with simulations allows us to avoid modeling the complex field evolution in boundary regions of the DC such as S-bends and tapers.

We measured the SR of the DCs across a broadband spectrum, but the shown curves in this work are all for $\lambda=1550\,nm$.
For the measurement, we used the amplified spontaneous emission (ASE) from an erbium-doped fiber amplifier (EDFA) and a superluminescent diode (SLED) as broadband sources at telecom wavelengths, and measured the output spectrum with a Yokogawa AQ6370D Optical Spectrum Analyser. 
The optical performance of the DCs was simulated using Ansys Lumerical's finite difference element (FDE) software. 

\section{Simulations with the Void}

\begin{figure}[!t]
    \centering
    \includegraphics[width=1\linewidth]{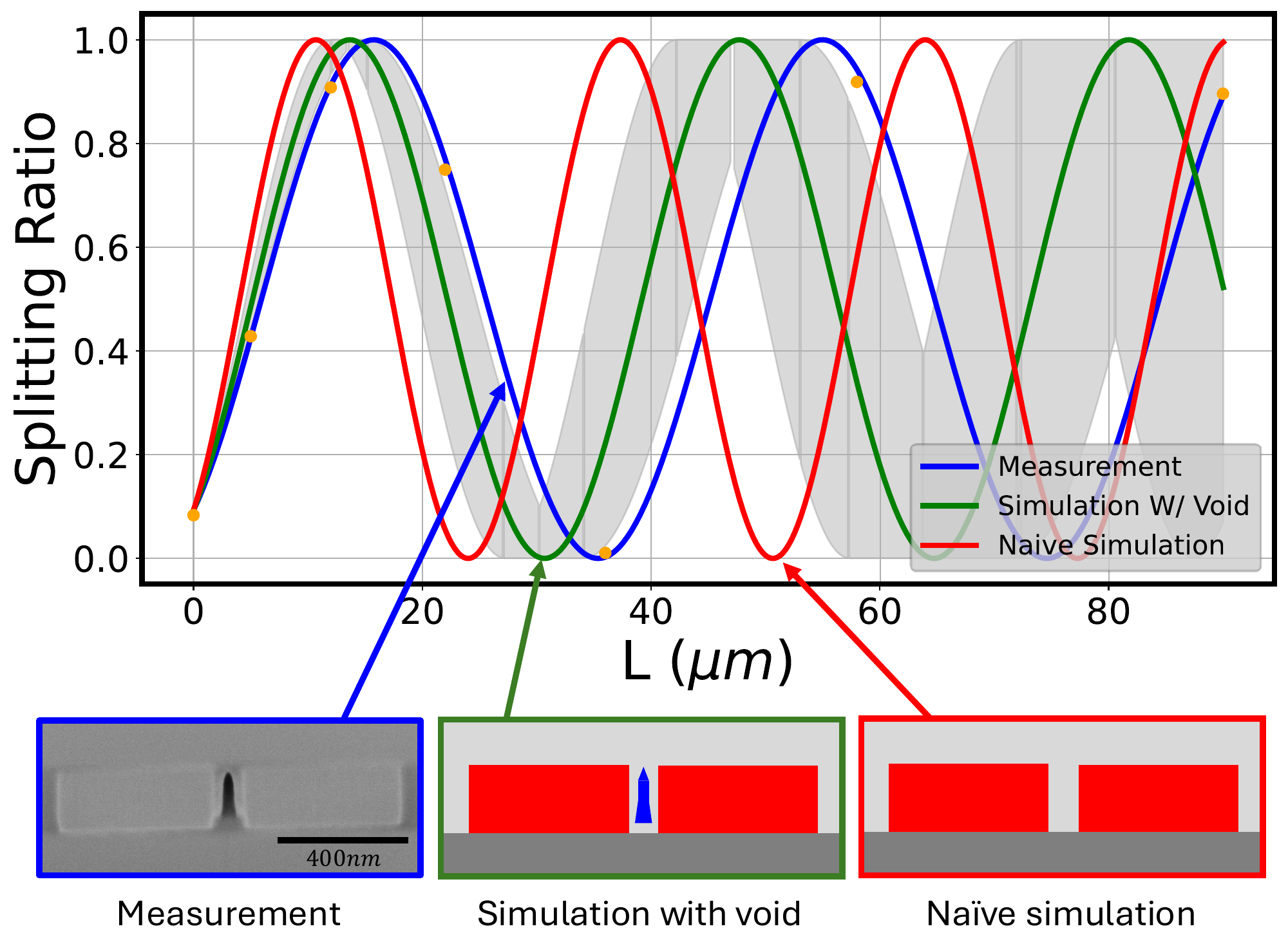}
    \caption{Comparison of experimental measurements with void model simulations. The DC's dimensions are $g_{ww}=87\,nm,\ w_1=w_2=500\,nm$, and the wavelength measured is $\lambda=1550\,nm$. Each measurement point is a different DC with the appropriate coupling region length. Here, the error bars are smaller than the data markers. The data was fitted to the curve given by Eq.~\ref{eq:SR}. The gray uncertainty bounds are the variation of the simulation under the geometry uncertainty provided by the manufacturer, as well as an uncertainty in the dimensions of the void. A typical coupling region length is $\sim15\,\mu m$. Over short length scales, the void seems to fix the discrepancy, while the naive simulation quickly diverges from the measurement. Over typical length scales, the measurement escapes the error bounds of the void simulation and diverges from the void simulation over longer scales.}
    \label{fig:void}
\end{figure}

\noindent As a first step, we imaged cross-sections of DCs using scanning and scanning transmission electron microscopes (SEM and STEM), as shown in Fig.~\ref{fig:dc_and_sem}. The cross-section images reveal that the cross-sectional geometry closely matches the planned design and falls well within the fabrication uncertainty specified by the manufacturer. However, there is a systematic reduction in $g_{cc}$ by $13 \text{nm}$, which has been accounted for in all simulations. The cross-sections also reveal a void in the cladding, that grows as the waveguides get closer. By imaging many cross-sections of DCs with different $g_{ww}$s, we fully characterized the void geometry, enabling its incorporation into the simulation.

In Fig.~\ref{fig:void} we show the result of the simulations using the extracted void geometry, along with the measurement and the naive simulation of a DC with $g_{ww}=87\,nm$. 
The simulations with the void show a decrease in the coupling, and over typical DC length scales of $\sim10\mu m$ the simulation with the void seems to predict the performance of the DC, compared to the naive simulation. However, over longer length scales it diverges from measurements and escapes the confidence bounds of the simulation.

We note this analysis was done for 13 different values of $g_{ww}$, where for each value the exact geometry of the void found from the SEM images was used. Fig.~\ref{fig:void} shows one instance of this analysis. Fig.~\ref{fig:coupling}a shows that the simulated coupling in the void model deviates from the measurements for all $g_{ww}$s.
We conclude that incorporating the void alone in the simulation does not fully resolve the discrepancy.

\section{The \textit{Effective Trench Medium Model}}

\begin{figure}[!t]
    \centering
    \includegraphics[width=1\linewidth]{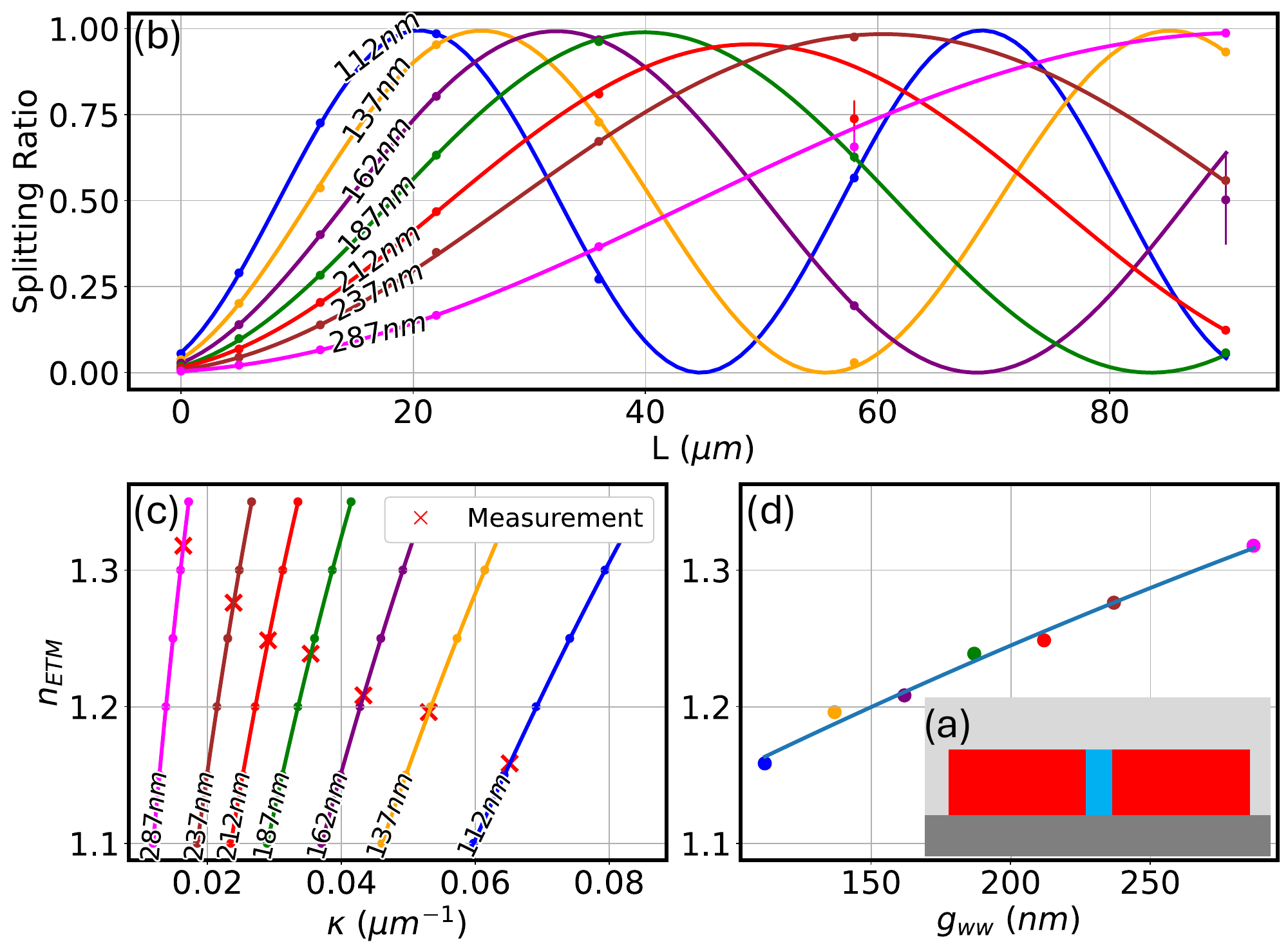}
    \caption{The calibration procedure of the Effective Trench Medium Model (ETMM). \textbf{(a)} The structure of the effective medium model. The effective medium (in light blue) occupies the volume between the waveguides (in red) while the rest of the cladding is assumed to be pure silica. The ETMM lumps the void and all other cladding inhomogeneities into an effective medium with a lower refractive index than the cladding. \textbf{(b)} Measured calibration data for the model. All measured DCs are symmetric with $w_1=w_2=500\,nm$, with different $g_{ww}$s, indicated on the curves. Each dataset was fitted to Eq.~\ref{eq:SR}. The coupling (oscillation frequency) was extracted for each $g_{ww}$. Where the error bar is not visible, the error is smaller than the data marker. \textbf{(c)} Finding the refractive index of the effective medium from simulation. Each curve is the result of a simulation of a DC with a different $g_{ww}$. Each point on a curve is the simulated coupling with a different value of $n_{ETM}$. The red Xs mark the measured coupling values from the calibration data. \textbf{(d)} The resulting model, the refractive index of the effective medium as a function of $g_{ww}$, with a quadratic fit.}
    \label{fig:eff med}
\end{figure}

\noindent Extending our effort to provide an accurate tool for simulative prediction of the performance of DCs, we have developed the ETMM. This model was inspired by previous work where low-density regions and void formation were reported in the $SiO_2$ cladding deposited by PECVD in high aspect ratio trenches \cite{schwartz1992gap, logan1989study, ohkubo2003fabrication, riley1989limitation}. In Fig.~\ref{fig:dc_and_sem}b, a noticeable difference in the cladding's color can be observed, with lighter-colored "wings" extending outward from the sides of the waveguides and a void forming at their intersection. It is worth noting that the shape of these "wings" are very similar to the density variations reported in previous studies \cite{logan1989study, schwartz1992gap}.

In this model, we assume that the trench between the waveguides has a uniform refractive index ($n_{ETM}$), which depends on the wall-to-wall gap of the DC alone, as illustrated in Fig.~\ref{fig:eff med}a.
The physical intuition to lump the variations in the refractive index between the waveguides is well-founded, as the wavelength of the propagating mode is $\sim 500\,nm$ ($\lambda = 1550\,nm$ in vacuum), while the spatial scale of the refractive index variations is much smaller, on the order of $\sim50\,nm$. As a result, the propagating mode perceives an effective medium with a uniform refractive index.

\begin{figure}[!t]
    \centering
    \includegraphics[width=1\linewidth]{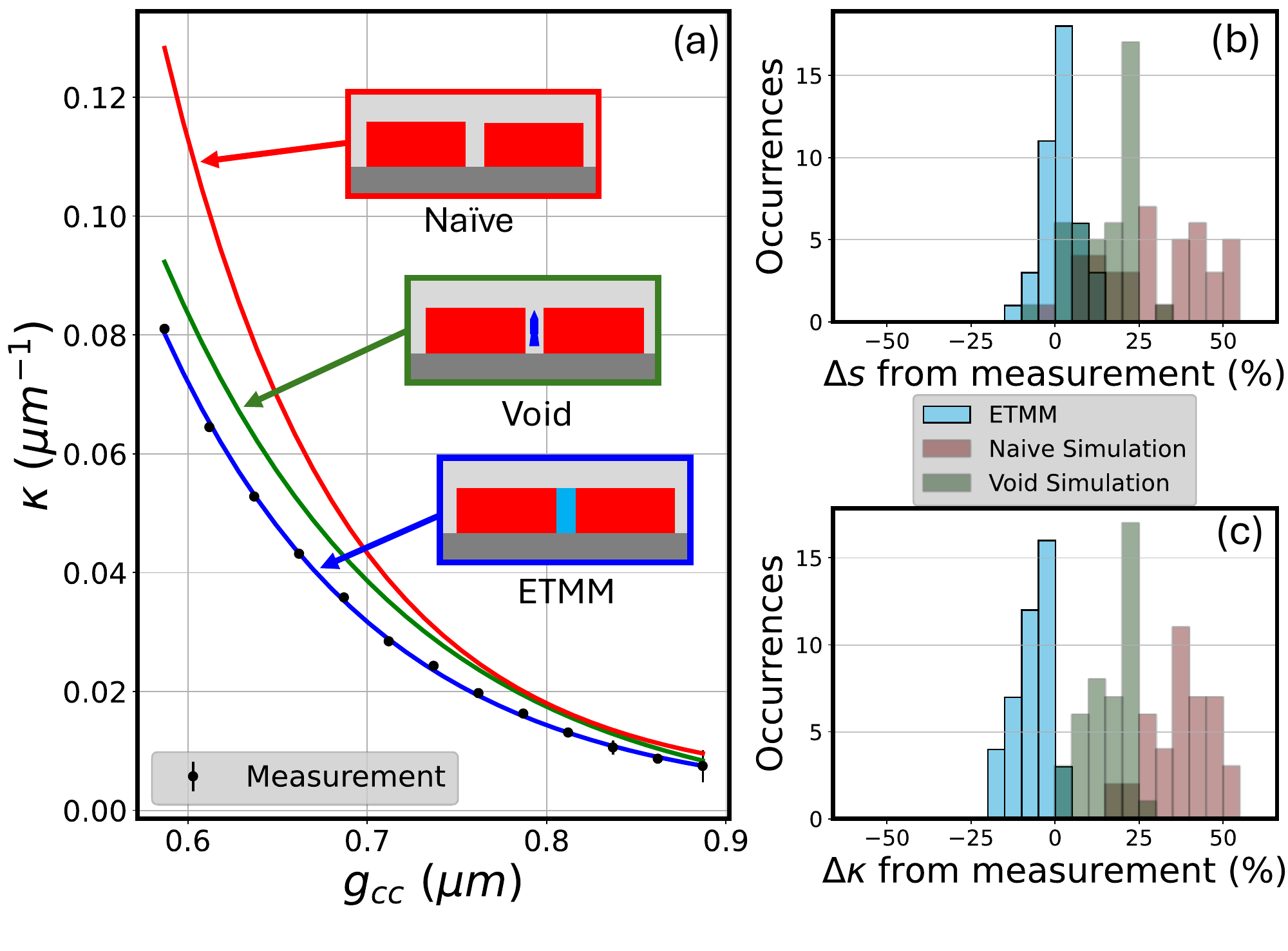}
    \caption{Comparison between models, and validation of the Effective Trench Medium Model (ETMM). \textbf{(a)} Coupling of a symmetric DC with $w_1=w_2=500\,nm$, measured and simulated by different methods. \textbf{(b+c)} Histograms of the differences in percentage of (b) the oscillation frequencies $s$ and (c) couplings $\kappa$ between the measured validation set and the different models.} 
    \label{fig:coupling}
\end{figure}

In Fig.~\ref{fig:eff med} we present the calibration procedure of the model. Fig.~\ref{fig:eff med}b shows a subset of the calibration dataset. These measurements are of symmetric DCs where the width of the waveguides is constant throughout the coupling region and S-bend ($w_1=w_2=500\,nm$), enabling the isolation of the coupling from the measurements using Eq.~\ref{eq:SR}. Each curve belongs to a DC with a different $g_{ww}$, and from each curve, the value of the coupling was extracted independently. We then determine the refractive index of the medium that, in simulation, produces the same coupling value as the measured one. Fig.~\ref{fig:eff med}c shows the simulated couplings for the DC geometries in the calibration set with several values of $n_{ETM}$. The measured coupling value for each geometry is marked on the curves, determining the refractive index of the effective medium. In Fig.~\ref{fig:eff med}d we display the final ETMM, namely we find $n_{ETM}$ as a function of $g_{ww}$, with a quadratic dependence, which was the minimal dependency that was required in order to fit the data. Namely, a linear model did not generalize well, and a cubic model over-fits approximately linear data. We observe that as $g_{ww}$ increases, the refractive index of the effective medium approaches that of the cladding. This is in tune with Fig.~\ref{fig:dc_and_sem} where the smaller $g_{ww}$ is, the larger the trench inhomogeneities become, and also in tune with Fig.~\ref{fig:coupling}a, where all the models' coupling curves converge for large values of $g_{cc}$.

To confirm the model's applicability beyond the calibration data, we have validated it on a separate dataset, consisting of asymmetric DCs with varying waveguide widths and $g_{ww}$. Each DC geometry was simulated naively, with a void and with the ETMM, which depends on $g_{ww}$ only. The oscillation frequency $s$ and coupling $\kappa$ were extracted from each simulation, and the difference from the measured validation set was calculated. Figs.~\ref{fig:coupling}b and \ref{fig:coupling}c show histograms, where each occurrence is the difference in $s$ and $\kappa$ respectively from the measured value, comparing all three simulation types. Both histograms validate the correctness of the ETMM and prove the remarkably better performance in recovering experimental results, both in terms of the mean values and spread. In Fig.~\ref{fig:coupling}b, the difference from the measurement of the inversion frequency $s$ of the ETMM is $1.4\pm5.3\%$ , an improvement by a factor of 11 from the simulation with the void with $15.2\pm8.3\%$ difference, and factor of 22 from the naive simulations with $30.2\pm14.8\%$ difference. 
Similarly, in Fig.~\ref{fig:coupling}c, the ETMM shows remarkably better predictions for the coupling $\kappa$ than the void and naive simulations, with a coupling difference of $-6.5\pm5.3\%$ for the ETMM, an improvement by a factor of 2.5 from $15.7\pm6.5\%$ for the void and factor of 6 from $37.5\pm9.1\%$ for the naive simulations.

The reason for the slight offset in the coupling validation histogram in Fig.~\ref{fig:coupling} has to do with adiabatic tapers in the asymmetric DCs. Due to the asymmetry in the waveguide width, both waveguides have short adiabatic tapers in the coupling region that cause a small shift in the propagation of the light that cannot be described simply by the coupling coefficient of the CMT, making the model in Eq.~\ref{eq:SR} less precise. Despite this shift, the ETMM still proves superior over the other simulations in the prediction of the coupling. To fully describe asymmetric DCs with tapers, finite difference time domain simulations incorporating propagation distance-dependent ETMM is needed. This is left for future work.

To extend the generalizability of the ETMM to other fabrication runs, we compared simulations with measurements for 62 DCs fabricated on an additional chip produced in a different fabrication cycle than the calibration chip. The results, summarized in Table \ref{tab:1}, indicate that the ETMM provides superior reproduction of the inversion frequency $s$ compared to the naive simulations. Although the ETMM achieved slightly less accuracy on this chip relative to the validation DCs on the original chip, which is expected given the known variations between fabrication runs, it still offers a significant improvement over the baseline simulations despite these variations. We note that identically cross-sectioned DCs on this chip were not placed randomly across it, but rather placed in series, leading to correlated shifts in the output, producing the large uncertainty in the reproduction of the inversion frequency. Despite these highly correlated errors, the ETMM still achieves to replicate the inversion frequency more accurately by a factor of 5 than the naive simulations.

Finally, the ETMM is compared with the measurements and the naive and void simulations in Fig.~\ref{fig:coupling}a, where the characteristic exponential coupling curves are shown. The naive simulation predicts dramatically different coupling values than the measurements, by as much as 45\%, in accord with the coupling loss reported by Tseng et al. \cite{Tseng:13}. The void model predicts different coupling values as well. Although closer, it deviates by up to 20\%. The model exactly predicts the measured coupling, as it was calibrated using these measurements.

\begin{table}[h]
    \centering

    \setlength{\tabcolsep}{4.5pt} 
    \renewcommand{\arraystretch}{1.2} 
    \begin{tabular}{lcccc}
        \toprule
        
        & $\Delta w(nm)$ & $g_{ww}(nm)$ & $\begin{matrix}
            \text{Cross-}\\ \text{section}\\ \text{designs}
        \end{matrix}$ &  $\begin{matrix}\text{Simulation vs.}\\ \text{Measurement of }s\end{matrix}$ \\

        \midrule

             \adjustbox{valign=c}{\shortstack{Calib-\\ration}}
      & $[0]$ 
      & $[87:13:387]$ 
      & 13 
      &  \\
                \midrule

        \adjustbox{valign=c}{\shortstack{Valid-\\ation:\\same\\chip}} & $[0:7:30]$ & $[130:6:180]$ & 42 & 
        $\begin{matrix}
            \text{ETMM: } 1.4 \pm 5.3\% \\ 
            \text{Naive: } 30.2 \pm 14.8\%
        \end{matrix}$ \\
        \midrule
        \adjustbox{valign=c}{\shortstack{Valid-\\ation:\\other\\chip}} & $[0:4:30]$ & $[100:12:210]$ & 62 &  
        $\begin{matrix}
            \text{ETMM: } 4.5 \pm 23.9\% \\ 
            \text{Naive: } 25.1 \pm 24.7\%
        \end{matrix}$ \\
        \bottomrule
    \end{tabular}
    \vspace{0.5em}
    \caption{\textnormal{Devices used for calibration and validation of the ETMM. The notation $[a:n:b]$ signifies $n$ values between $a$ and $b$. The nominal $g_{cc}$ for all validation devices is $650nm$. The second validation set contains devices with identical $\Delta w$ and $g_{ww}$ but different average waveguide widths, thus the set does not contain all possible parameter combinations and has some repetitions unlike the first validation set.}}

    \label{tab:1}
\end{table}

\section{Discussion}


\noindent An underlying assumption in the ETMM is the presence of density variations in the cladding. Although we could not measure it directly, the SEM and STEM images presented in  Fig.~\ref{fig:dc_and_sem} show patterns in the cladding consistent with previous works, and more importantly, the void alone is insufficient in explaining the discrepancy. We have shown that by including the void only in the simulations moves them in the right direction yet they still fall short. By excluding all other possible geometrical discrepancies together with previous reports of low-density regions in high aspect ratio trenches using PECVD \cite{schwartz1992gap, logan1989study, ohkubo2003fabrication, riley1989limitation}, provide a strong motivation for our model.


Although the PECVD process is widely used throughout the industry, some platforms have other approaches for applying cladding, or avoid applying cladding altogether thereby eliminating cladding inhomogeneities.
However, without a cladding layer, it is impossible to integrate critical active control devices onto the chip such as heaters and capacitors. Also, the nanoscale devices will be in direct contact with the environment, exposing them to contamination which can cause unpredictable outcomes.
Another method of applying an oxide cladding is a variant of PECVD called high-density plasma (HDP). This process has superior gap-filling characteristics but comes at the cost of damaging ion bombardment of the silicon devices \cite{pai1996high}. An additional proposed method to eliminate the cladding density variations is Atomic Layer Deposition (ALD) \cite{choi2022bottom}. However, it can only be used on individual dies and not on entire wafers, and is currently an extremely slow and expensive process.

Our findings also have major implications for integrated photonic design as a whole, specifically on optimization-based design. Precise simulation is a keystone in optimization-based design, which relies on highly accurate data for the optimization process. These processes require huge amounts of data, which cannot be gathered experimentally, and must be simulated reliably instead. Our work enables the creation of the necessary accurate data, opening the floodgates of reliable optimization-based design of PIC devices. Furthermore, we present a structured calibration framework that details the steps required to measure a representative set of directional couplers with varying gaps and extract the corresponding effective refractive index. This framework not only underpins the success of our current approach on the ANT platform but also facilitates systematic adaptation to other fabrication environments with minimal additional experimental data. While our present work focuses on directional couplers, we recognize that extending the ETMM to more complex devices, such as MMIs and grating couplers, will require additional data to accurately capture their unique dynamics affected by the oxide-density variations. Nevertheless, the calibration framework established herein provides a promising foundation for such adaptations.

\section{Conclusion}

\noindent As integrated photonic platforms continue to evolve, directional couplers remain central to realizing scalable and efficient PICs. We have shown that in DCs covered by oxide cladding, voids and density variations are formed in the cladding between the waveguides and are dependent on the wall-to-wall gap between the waveguides. These inhomogeneities, unaccounted for, are the cause for the dramatic discrepancy between measurement and simulation. We have quantified their effect on the dynamics of light in DCs, and have shown that by compensating for it in simulations with a novel ETMM, measurements of DCs can be predicted with unprecedented accuracy. Our findings pave the way for accurate and faithful simulations of PICs, a crucial step in advancing the field of integrated photonics.

\section*{Acknowledgment}
\noindent The authors express sincere thanks to Dr. Alex Lahav of Tel Aviv University for help in the operation of the electron microscopes and related guidance. The authors also sincerely thank Applied Nanotools Inc. (ANT) and Cameron Horvath for their continuous support and device fabrication.

\section*{Funding}
\noindent Israel Science Foundation (2312/21) and Israel Science Foundation Quantum (3427/21).

\bibliographystyle{IEEEtran}
\bibliography{references}

\end{document}